# Generating functions partitioning algorithm for computing power indices in weighted voting games

**Pessimistic computational complexity $O(n\,2^{n/2})$**

**Psuedopolynomial computational complexity $O(nq)$**

Bartosz Meglicki

**Abstract:**

In this paper new approach to calculating power indices is described. The problem complexity class is #P-complete. Constructed algorithm is a mix of ideas of two algorithms: Klinz & Woeginger partitioning algorithm and Mann & Shapley generating functions algorithm. Time and space complexities of the algorithm are analysed and compared with other known algorithms for the problem. Constructed algorithm has pessimistic time complexity $O(n2^{n/2})$ and pseudopolynomial complexity $O(nq)$, where $q$ is quota of the voting game. This paper also solves open problem stated in [2] - existence of the algorithm for calculating Banzhaf power indices of all players with time complexity lower than $O(n^2 2^{n/2})$. Not only is the answer positive but this can be done keeping the pseudopolynomial complexity of generating functions algorithm in case weights are integers. New open problems are stated.

## Table of contents

## 1 Introduction

Weighed voting games are mathematical models of voting bodies in which each player has certain number of votes. In mathematical model each voter is assigned weight and votes for or against a decision. The decision is accepted if sum of weights of players voting in favour of decision is greater or equal to fixed quota.

It can be easily proven that the power of a player in a voting body isn't equal to his weight (absolute or relative). Power indices are objective way of assessing the power of a player in voting body. Banzhaf power index is one of the most commonly accepted among voting power theorists.

In general, when no assumptions about players weights can be made, calculation of Banzhaf indices of players is a difficult task. In this paper effective algorithm for calculating Banzhaf power indices is devised.

## 2 Preliminaries

**Defs 1** A *simple voting game* is a pair $G=(P,v)$ where:

$P=\{1,2,3,...,n\}$ - finite non-empty set of players

$v:2^P \to \{0,1\}$ - function satisfaying $v(\emptyset)=0$, $v(P)=1$, $S \subseteq T \Rightarrow v(S) \leq v(T)$

Coalition $C \subseteq P$ is *winning* if $v(C)=1$, *losing* if $v(C)=0$

A simple game is *proper* if $\forall C \subseteq P \ v(C)+v(P\setminus C) \leq 1$

**Defs 2** Symbol $[q;w_1,w_2,...,w_n]$ represents a simple voting game:

$$G=(P,v) \quad P=\{1,2,3,...,n\}$$
$$v(C)=\begin{cases} 1 & for \ w(C) \geq q \\ 0 & for \ w(C) < q \end{cases} \quad C \subseteq P, w(C)=\sum_{i \in C} w_i \qquad (1)$$

Such game is called *weighted voting game*.

Weighted voting game is proper if $q > w(P)/2$

**Defs 3** A player *p* is critical in coalition *C* if:

$$v(C)=1 \wedge v(C\setminus\{p\})=0 \qquad (2)$$

For player *p*, we denote the number of coalitions in which *p* is critical in game *G* as:

$$\eta_p(G)=|\{C \subseteq P : v(C)=1 \wedge v(C\setminus\{p\})=0\}| \qquad (3)$$

**Def. 4** The *Banzhaf index* of player *p* in game *G* is:

$$\beta_p = \eta_p(G) / \sum_{i \in P} \eta_i(G) \qquad (4)$$

**Def. 5** *Probabilistic Banzhaf index* of player *p* in game *G* is:

$$\beta'_p = \eta_p(G)/2^{n-1} \qquad (5)$$

## 3 Problem

For a weighted voting game $[q;w_1,w_2,...,w_n]$, we are interested in calculating Banzhaf indices of all players. Even if we only want to calculate probabilistic Banzhaf index of the biggest player the problem is NP-hard [10].

The problem of calculating Banzhaf indices of players is #P-complete [14]. It means that it is as hard as any counting problem in NP (examples: #SAT, Permanent, #HAMILTON PATH, the counting problem for the knapsack problem). The counting problem is at least as hard as the corresponding NP problem. For example if we could solve #SAT problem, we can solve decision SAT problem by asking a question - "is the number of solutions of #SAT problem greater than zero?".

## 4 Known solutions

There are exact and approximate algorithms for the problem of calculating Banzhaf power index. Here only exact algorithms are considered. The basic naive algorithm enumerates through all coalitions and has time complexity $O(n2^n)$ and memory complexity $O(n)$.

Klinz and Woeginger in [7] presented algorithm which calculates Banzhaf indices of all players in time $O(n^2 2^{n/2})=O(n^2 \sqrt{2}^n)=O(n^2 1.41^n)$. The probabilistic Banzhaf index of a single player can be calculated in time bounded by $O(n 2^{n/2})$. The memory complexity of the algorithm is $O(2^{n/2})$. The idea of the algorithm is to partition the set of players into two disjoint subsets, find all possible coalitions inside those subsets, sort coalitions by weights and apply technical algorithm developed by Horowitz and Sahni in [6].

Mann and Shapley in [9] developed dynamic programming algorithm based on generating functions for calculating Shapley-Shubik index of power. Brams and Affuso in [5] modified the algorithm to calculate Banzhaf index. The generating functions algorithm is pseudopolynomial. If the numbers in the instance of the problem are small, the algorithm behaves as polynomial algorithm. In general, if we allow any input, the algorithm has exponential pessimistic complexity.

Bilbao at all in [4] show generating functions algorithm for calculating Banzhaf power index for all players in time bounded by $O(n^2 C)$. *C* is the number of non-zero generating function coefficients and is equal to the number of distinct sums of weights that can be obtained by forming coalitions. Also $C \leq min(2^n, \sum_{i \in P} w_i)$ and it's possible to implement the algorithm in a way that $C \leq min(2^n, q)$.

Uno [17] shows table generating functions implementation with pseudopolynomial complexity $O(nq)$ for calculating indices of all players. The memory complexity of the algorithm is $O(q)$.

Owen in [11,12] developed multilinear extension algorithm for voting games. The algorithm is exact and has exponential time complexity. With the use of central limit theorem algorithm is the basis of most approximation algorithms.

Matsui in [10] show recursive algorithm enumerating through all minimal winning coalitions. Algorithm can calculate Banzhaf index of single player in time bounded by $O(n^2 |minW|)$, where *minW* is the family of minimal winning coalitions (coalition is minimal if it's every player is critical). The memory complexity of the algorithm is $O(n^2)$.

## 5 Outline of the solution

Algorithm presented in this paper is a mix of two ideas: partitioning approach by Horowitz and Sahni [6] and generating functions by Mann and Shapley [9]. The partitioning approach was adapted for weighted voting games by Klinz and Woeginger [7].

We partition the set of all players into two subsets with $\lceil n/2 \rceil$ and $\lfloor n/2 \rfloor$ players. In each of the subsets independently we obtain generating function. Next we remove single player from one of the subsets, dividing generating function adequately. To solve our original problem we modify and use algorithm for technical problem by Horowitz and Sahni [6]. In both subsets generating functions have at most $2^{n/2}$ non-zero coefficients. Horowitz and Sahni approach, for sorted data, allows us to solve original problem in time bounded by $O(N+M)$, where *N* and *M* are the numbers of elements in the subsets (in our case number of non-zero generating function coefficients in both subsets)

Algorithm combines properties of Klinz and Woeginger partitioning algorithm and generating functions algorithm. For calculating indices of all players, it has pessimistic time complexity $O(n 2^{n/2})$ and pseudopolynomial time complexity $O(nq)$, where *q* is the quota of weighted voting game. Algorithm has psedopolynomial complexity only if weights are integers.

# 6 Technical problem

After [6,7], we will construct effective algorithm for technical problem:

**Input:**

$x_1 \leq x_2 \leq ... \leq x_M$  $x_i \geq 0 \wedge x_i \in \mathbb{Z}$ - M nonnegative integers sorted ascending

$y_1 \leq y_2 \leq ... \leq y_N$  $y_i \geq 0 \wedge y_i \in \mathbb{Z}$ - N nonnegative integers sorted ascending

$a_1, a_2, ..., a_M$  $a_i \geq 0 \wedge a_i \in \mathbb{Z}$ - M nonnegative integers

$b_1, b_2, ..., b_N$  $b_i \geq 0 \wedge b_i \in \mathbb{Z}$ - N nonnegative integers

$L, U \in \mathbb{Z}$  $L \leq U$ - integer lower and upper bounds

**Problem:**

What is the sum of all products $a_i b_j$, taken over all indices pairs $i,j$  $1 \leq i \leq M$  $1 \leq j \leq N$ that satisfy $L \leq x_i + y_j \leq U$? Formally we ask for X:

$$X = \sum_{i=1}^{M} \sum_{j=1}^{N} g(x_i + y_j) a_i b_j$$

$$g : \mathbb{Z} \to \mathbb{Z}  \quad g(n) = \begin{cases} 1 & L \leq n \leq U \\ 0 & else \end{cases}$$

(6)

**Theorem 1** Technical problem can be solved in time O(N+M) and memory O(N+M)

**Proof** We will construct algorithm.

We add to the input data dummy values $y_0 = -\infty$ and $y_{M+1} = \infty$, which will serve as sentinels. It implies:

$$\forall i (x_i + y_0 < L \leq U \ \wedge \ x_i + y_{M+1} > U \geq L)$$

(7)

For $1 \leq i \leq M$ we compute the smallest *l=l(i)*, such that $x_i + y_l \geq L$. From 7 *l(i)* is properly defined for each *i*. Since the numbers $x_i$ and $y_i$ are sorted ascending, for each *i* $l(i+1) \leq l(i)$ is satisfied.

l(1) is computed by searching from $y_j = y_{N+1}$ to $y_j = y_0$, until $x_1 + y_j < L$ is satisfied. Then we set l(1)=j+1.

Since $l(i+1) \leq l(i)$ is satisfied, for $i > 1$ we search for l(i) from $y_j = y_{l(i-1)}$ to $y_l = y_0$ until $x_i + y_j < L$ is satisfied. Then we set $l(i) = j+1$.

The number of search steps for calculating l(1) is N+2-l(1). For $i > 1$ we start searching from l(i-1) and finish on l(i) and thus the number of search steps for l(i) is $l(i-1) - l(i) + 2$. The overall number of steps for calculating l(i) for $1 \leq i \leq M$ is:

$$N - l(1) + 2 + \sum_{i=2}^{M} (l(i-1) - l(i) + 2) = N - l(M) + 2M = O(M+N)$$

(8)

Next, for $1 \leq i \leq M$ we compute the largest *u=u(i)*, such that $x_i + y_u \leq U$. From 7 *u(i)* is properly defined. For each *i* it is satisfied that $u(i+1) \leq u(i)$. We can compute *u(i)* values in a procedure symmetric to this for l(i).

For any index *i* we have:

$$L \leq x_i + y_j \leq U \Leftrightarrow l(i) \leq j \leq u(i) \quad (9)$$

Next for $0 \leq m \leq M$ we compute and store the values:

$$F(m) = \sum_{i=1}^{m} b_i \quad (10)$$

This can be done in $O(M)$ time by adding $b_m$ to the sum in each step.

Finally we can find the answer to technical problem.

Since, for a fixed *i*, the condition $L \leq x_i + y_j \leq U$ is equivalent to $l(i) \leq j \leq u(i)$ the $g(x_i + y_j)$ function value from 6 is nonzero only if $l(i) \leq j \leq u(i)$ holds. Therefore we can obtain X:

$$X = \sum_{i=1}^{M} \sum_{j=1}^{N} g(x_i + y_j) a_i b_j = \sum_{i=1}^{M} \sum_{j=l(i)}^{u(i)} a_i b_j = \sum_{i=1}^{M} a_i \sum_{j=l(i)}^{u(i)} b_j = \sum_{i=1}^{M} a_i [F(u(i)) - F(l(i)-1)] \quad (11)$$

Hence by using formerly computed *F(m)* values, 11 can be calculated in O(M) time.

In the whole algorithm we use additional memory to store the values of F(m), l(i) and u(i), $0 \leq i, m \leq M$. Therefore memory complexity of the algorithm is O(M+N).

Calculating *l(i)* and *u(i)* for $1 \leq i \leq M$ needs O(M+N) operations. Calculating and storage of F values for $0 \leq m \leq M$ needs O(M) operations. Calculating X value, as in 11, needs O(M) operations. Thus the algorithm complexity is O(M+N) which ends the proof.

## 7 Generating functions

Generating function of a number sequence $(a_0, a_1, a_2, ...)$ is a formal power series:

$$G(x) = \sum_{k=0}^{\infty} a_k x^k \quad (12)$$

This power series is called formal since we are not interested in it's value for particular *x* and convergence problems. What we are interested in are it's coefficients $a_k$. We encode our problem in such a way that the coefficient corresponding $x^k$ is equal to the number of coalitions, possible to form, with weight equal to *k*.

Let $(P, v)$ be a voting game represented by $[q; w_1, w_2, ..., w_n]$. We will write *w* for the sum of weights of all players, *w=w(P)*. The biggest coalition (of all players) in the game has weight sum *w*. As a consequence all the terms in the sequence $\{a_k\}_{k \geq 0}$ from $k > w$ are equal to zero. Hence we can write:

$$G(x) = \sum_{k=0}^{w} a_k x^k \quad (13)$$

The generating function G(x) can be obtained by using elementary operations on polynomials.

**Theorem 2** Let $(P, v) = [q; w_1, w_2, ..., w_n]$ be a voting game. The generating function of a sequence $(a_0, a_1, a_2, ...)$, where $a_k$ is equal to the number of coalitions with weight sum *k* is:

$$G(x) = \prod_{i=1}^{n} (1 + x^{w_i}) \quad (14)$$

**Proof.**

$$\prod_{i=1}^{n}(1+x^{w_i})=(1+x^{w_1})(1+x^{w_2})...(1+x^{w_n})=\sum_{S\subseteq P}\prod_{i\in S}x^{w_i}=\sum_{S\subseteq P}x^{\sum_{i\in S}w_i}=\sum_{k=0}^{\infty}a_k x^k \quad (15)$$

The product of the factors on the left side of equation 15 can be seen as the sum taken over all possible choices of parentheses. In a single choice we have S from P possible parentheses. From the chosen parentheses we multiply corresponding $x^{w_i}$, from those that are not chosen we multiply by 1. Then we notice that:

$$\prod_{i\in C}x^{w_i}=x^{\sum_{i\in C}w_i}$$

Summing up such expressions we get a polynomial (since from some *k* all the terms are equal to zero). Choosing parentheses corresponds to choosing players to coalition. Multiplication $x^i x^j = x^{i+j}$ corresponds joining coalition with weight sum *i* and coalition with weight sum *j*. By summing up all the monomials with the same power $x^k$, we count the number of coalitions with that sum *k*.

Let *j* be a step in multiplying terms of 14 by $(1+x^{w_j})$. We will analyse the relation between coefficients $a_k^{(j)}$ and $a_k^{(j-1)}$. Since there is only one way to obtain a coalition with weights sum zero (the empty coalition), for each *j* we have $a_0^{(j)}=1$. Moreover $a_k^{(0)}=0$ for $1\le k\le n$. When we are multiplying polynomial from step *(j-1)* in step *j*, we have:

$$(1+a_1^{(j-1)}x+a_2^{(j-1)}x^2+...+a_w^{(j-1)}x^w)(1+x^{w_j}) \quad (16)$$

From multiplying by 1 we get the same polynomial as in step *(j-1)*. By multiplying by $x^{w_j}$ we modify the coefficients $a_k^{(j-1)}$ in the following way: $a_k^{(j)}=a_k^{(j-1)}+a_{k-w_j}^{(j-1)}$ for $k\ge w_j$. As a consequence we get recurrence:

$$a_k^{(j)}=\begin{cases}a_k^{(j-1)}+a_{k-w_j}^{(j-1)} & k\ge w_j \\ a_k^{(j-1)} & k<w_j\end{cases} \quad (17)$$

After applying 17 *n* times, in step *j* for $0\le k=w(P)$, we get coefficients $a_k$ for $0\le k\le w(P)$, equal to the number of coalitions with weights sum *k*. In the next theorem we will show how to calculate coefficients $a_k$ in time bounded by the number of players in game and number of non-zero generating function coefficients.

**Theorem 3** Let *C* be *number* of non-zero coefficients in generating function 14. The values of coefficients $a_k$ for $0\le k\le w$ can be calculated in time $O(nC)$ and memory $O(C)$.

**Proof** We will construct algorithm.

We will be putting non-zero GF coefficients on the list along with the power of *x* they correspond to. First we put on the list a pair (0,1) - the empty coalition can be formed in only one way.

By multiplying factors $(1+x^k)$ in 14, from multiplication by 1 we get the same polynomial and from multiplication by $x^k$ we get polynomial with coefficients increased by *k*. In each step we will enumerate through the list with coefficients and corresponding powers of *x*. On auxiliary list we will be putting elements corresponding to the powers of *x* increased by *k*.

Since elements on both lists are sorted ascending we can merge lists (it corresponds to summing the polynomials), using procedure similar to the one used in classical MergeSort algorithm. The only difference is in case the elements on both lists correspond to the same power of *x*. In such case we sum up values of coefficients equal to the number of ways in which coalitions can be formed and we put on the output list single element.

### Algorithm 1

```
Input: voting game [q, w₁, w₂,...,ₑₙ]
Output: list with non-zero GF 14 coefficients along with corresponding powers of
x

ListElement
{
   int XPower; //the power of x in monomial from generating function
   int Count;  //the a coeffict in monomial from generating function
   ListElement Next; //the next element on the list
}

 1: List GeneratingFunction(n, [q, w₁, w₂,...,wₙ])
 2: {
 3:    List A,B;
 4:    A.Push( (0,0) );
 5:    for(i=0;i<n;i++)
 6:    {
 7:       ListElement temp=A.First;
 8:       while(temp!=A.ListEnd)
 9:       {
10:          B.Push( (temp.XPower+i,temp.Count) );
11:          temp=temp.Next;
12:       }
13:       A=Merge(A,B);
14:    }
15:    return A;
16: }
17: List Merge(A,B)
18: {
19:    List R;
20:    R.Push( (0,0) );
21:    while( A.NotEmpty() or B.NotEmpty() )
22:    {
23:       if( A.NotEmpty() and B.NotEmpty() )
24:          if(A.First.XPower==R.Last.XPower)
25:             R.Last.Count+=A.First.Count; A.RemoveFirst();
26:          else if(B.First.XPower==R.Last.XPower)
27:             R.Last.Count+=B.First.Count, B.RemoveFirst();
28:          else if(A.First()<B.First() )
29:             R.Append(A.First), A.RemoveFirst();
30:          else
31:             R.Append(B.First), B.RemoveFirst(), continue;
32:       else if( A.NotEmpty() ) //B is empty
33:          if(A.First.XPower==R.Last.XPower)
34:             R.Last.Count+=A.First.Count; A.RemoveFirst();
35:          R.Append(A), A.Clear();
36:       else if( B.NotEmpty() ) //A is empty
37:          if(B.First.XPower==R.Last.XPower)
38:             R.Last.Count+=B.First.Count; B.RemoveFirst();
39:          R.Append(B), B.Clear();
40:    }
41: }
```

**Algorithm complexity:**

The loop in lines 5-14 will execute *n* times. In each step we go through the list with length bounded by *C* (the number of non-zero GF coefficients, defined in 14). The auxiliary list has the same amount of elements as list A and thus it's length is also bounded by *C*. The Merge function of classic MergeSort algorithm has complexity O(m+n) where *m* and *n* are lengths of the merged lists. In our case Merge is O(C), hence the time complexity of the whole algorithm is O(nC). As both lists and output list in each step are bounded by C we can never keep more than 3C elements on the lists, thus memory complexity is O(C).

To solve our problem we also need generating function of sequence $\{c_{k,l}\}_{k\geq 0}$. We encode the problem in such a way, that for a given player $l$ $c_{k,l}$ corresponds to the number of coalitions with weights sum $k$ but without player $l$.

For a player $l$, the generating function of $\{c_{k,l}\}_{k\geq 0}$ can be obtained by dividing the generating function defined in 14 by $(1+x^{w_l})$:

$$H(x) = \sum_{k=0}^{w} c_{k,l} x^k = G(x)/(1+x^{w_l}) = [\prod_{i=1}^{n}(1+x^{w_i})]/(1+x^{w_l}) \tag{18}$$

It holds:

$$(1 + c_{1,l} x + c_{2,l} x^2 + \ldots + c_{v,l} x^v)(1+x^{w_l}) = 1 + a_1 x + a_2 x^2 + \ldots + a_w x^w \tag{19}$$

Comparing 19 with 16 we notice that the dependence is the same as in 17, hence:

$$a_k = \begin{cases} c_{k,l} + c_{k-w_j,l} & k \geq w_l \\ c_{k,l} & k < w_l \end{cases} \tag{20}$$

By transforming 20 to obtain $c_{k,l}$ we get:

$$c_{k,l}^{(j)} = \begin{cases} a_k - c_{k-w_l,l}^{(j-1)} & k \geq w_l \\ a_k & k < w_l \end{cases} \tag{21}$$

21 should be applied from the smallest to the largest $k$, that is $k = 0, 1, \ldots, w$.

Notation $(j)$ in $c_{k,l}^{(j)}$, means the step incremented along with incrementation of $k$.

We will renumerate 21:

$$c_{k+w_l,l}^{(j)} = \begin{cases} a_{k+w_l} - c_{k,l}^{(j-1)} & k \geq 0 \\ a_{k+w_l} & k < 0 \end{cases} \tag{22}$$

In the next theorem we will show how to calculate $c_{k,l}$ in time and memory bounded by the number or players and non-zero coefficients of generating function.

**Theorem 4** Let's assume that we have non-zero $a_k$ coefficients of GF defined in 13. Let $C$ be the number of those coefficients. The values of coefficients $c_{k,l}$ for $0 \leq k \leq w$ and given $l$ can be calculated in $O(C)$ time and memory.

**Proof** We will construct algorithm.

Let A be a given list with pairs of coefficients $a_k$ and corresponding powers $k$. We will be moving through the list A and putting elements on the auxiliary list B. On the list B we put values by which $a_{k+w_l}$ should be decreased according to 22. Upon moving to each $a_k$ from the list A we will be checking what element lies at the beginning of the list B. In case the power of x of element in B is lower than $k$ of $a_k$, we remove that element from B. We keep on removing such elements from B until the power of x in the B element is greater or equal $k$ of current $a_k$. If the power of x of element in B corresponds to $k$ of $a_k$, we decrease $a_k$ value adequately and remove this element from list B.

## Algorithm 2

```
Input: list of non-zero GF 14 coefficients along with the powers of x, weight of
player l
Output: list of GF 18 coefficients(some of them may equal zero)

ListElement
{
   int XPower; //the power of x in monomial from GF
   int Count;  //coefficent a in monomial from GF
   ListElement Next; //next element on the list
}

1:  List GFDivide(List A,int w)
2:  {
3:     List B;
4:
5:     ListElement temp=A.First;
6:
7:     while(temp!=A.ListEnd)
8:     {
9:        while(B.NotEmpty() && B.First.XPower<temp.XPower)
10:          B.RemoveFirst();
11:
12:       if(B.NotEmpty() && B.First.XPower==temp.XPower)
13:          temp.Count-=B.First.Count, B.RemoveFirst();
14:       B.Push( (temp.XPower+w,temp.Count) );
15:       temp=temp.Next;
16:    }
17:    B.Clear();
18:    return A;
19: }
```

**Algorithm complexity:**

There are C elements on list A. Therefore the loop in lines 7-16 executes C times. For each element from A we put corresponding element to B. Each element in B is considered only once in lines 9-13 and then removed. Thus both time and space complexity of the algorithm are O(C).

It is worth noting that this algorithm leaves coefficients $c_{k,l}$ equal to zero on the output list if $a_k$ was non-zero.

# 8  Main result

According to formula 2, player *p* is critical in coalition *C* if:

$$w(C) \geq q \wedge w(C) - w_p \leq q - 1 \tag{23}$$

By subtracting player's *p* weight in inequalities we get interval in which coalition (without *p*) weight sum has to be in order for the player *p* to be critical.

$$q - w_p \leq w(C) - w_p \leq q - 1 \tag{24}$$

The classic generating functions algorithm [9], uses inequalities from 24 and generating function 18 to calculate the number of critical coalitions for player *p*. The $c_{k,p}$ coefficients of the generating function 18 are equal to the number of coalitions without player *p* with weight sum *k*. Hence, the number of critical coalitions for player *p* can be calculated as:

$$\eta_p(z) = \sum_{i=q-w_p}^{q-1} c_{i,p} \tag{25}$$

Our approach will be different. We will construct the answer to the problem by partitioning the set of players into two subsets. The algorithm is based on the observation described in next paragraph.

Let S and T be two disjoint subsets of set of players, $S \cap T = \emptyset$. Assume that we know that inside of subset S the number of coalitions with weights sum *x* is equal to *a* and inside of set T the number of coalitions with weight sum *y* is equal to *b*. Then the number of coalitions with weights sum *x+y* inside set $S \cup T$ is equal to *ab* (perhaps it is also possible to get sum *x+y* by combining other sums of coalitions but it is irrelevant now)

Having generating functions from sets S and T we could answer the question: "What is the number of coalitions with sum *x+y* inside set $S \cup T$?" We could, for instance, multiply the generating functions.

Let $z = (P, v) = [q, w_1, w_2, \ldots, w_n]$ be a voting game. We partition the set of players *P* into two disjoint subsets *A* and *B*. Then we remove player *p* from set *A*. Let $G_A(x)$ be a generating function obtained from set $A \setminus \{p\}$ and $G_B(x)$ from set B.

$$G_A(x) = \sum_{i=1}^{\infty} a_k x^k \quad G_B(x) = \sum_{i=1}^{\infty} b_k x^k$$

The $a_k$ coefficients are equal to the number of coalitions with weights sum k in $A \setminus \{p\}$ and $b_k$ in set $B$.

The number of critical coalitions for player *p* could be calculated as:

$$\eta_p(z) = \sum_{i=1}^{\infty} \sum_{j=1}^{\infty} g(i+j) a_i b_j$$

$$g(n) = \begin{cases} 1 & q - w_p \leq n \leq q - 1 \\ 0 & else \end{cases} \tag{26}$$

All the coefficients $a_i$ and $b_j$ beginning from some *i* and *j* are equal to 0. Hence we could write the sum above as:

$$\eta_p(z) = \sum_{i=1}^{w(A)} \sum_{j=1}^{w(B)} g(i+j) a_i b_j \tag{27}$$

Notice that 27 is the answer to our technical problem from section 6 (compare 27 with 6). We will construct algorithm based on those observations.

**Algorithm:**

Let $D=(P,v)=[q, w_1, w_2, ..., w_n]$

1. We partition the set of players *P* into two subsets *A* i *B* with the following cardinalities:

$$|A|=\lceil \frac{|P|}{2}\rceil \text{ and } |B|=\lfloor \frac{|P|}{2}\rfloor \tag{28}$$

2. We obtain generating functions $G_A(x)$ and $G_B(x)$ using algorithm 1 for voting games:

$$D_A=(A,v)=[q, w_1, w_2, ..., w_{|A|}] \text{ and } D_B=(B,v)=[q, w_{|A|+1}, w_{|A|+2}, ..., w_n] \tag{29}$$

Let $K_A$ and $K_B$ be the lists of generating function coefficients obtained from A and B.

3. We will calculate the number of coalitions with critical player *p* from set *A*. The case when player p is from set *B* is symmetric. We divide generating function $G_A(x)$ by $(1+x^{w_p})$ using algorithm 2 for a copy of list $K_A$ and $w=w_p$. Let $K_C$ be the resulting coefficients list.

Let $\alpha$ be the length of list $K_C$ and it's content be:

$$((k_1, a_{k_1}), (k_2, a_{k_2}), ..., (k_\alpha, a_{k_\alpha}))$$

, where $a_0 x^0$ from the generating function corresponds to $(k_1, a_{k_1})$ and the following elements from the list correspond to following monomials of generating function.

Let $\beta$ be the length of list $K_B$ and it's content be:

$$((l_1, b_{l_1}), (l_2, b_{l_2}), ..., (l_\beta, b_{l_\beta}))$$

, where following elements from the list correspond to non-zero monomials of generating function $G_B(x)$.

4. We solve the technical problem from section 6 using input:

| Technical problem notation | Supplied data |
|---|---|
| $M$ | $\alpha$ |
| $N$ | $\beta$ |
| $(x_1, x_2, ..., x_M)$ | $(k_1, k_2, ..., k_\alpha)$ |
| $(y_1, y_2, ..., y_M)$ | $(l_1, l_2, ..., l_\beta)$ |
| $(a_1, a_2, ..., a_M)$ | $(a_{k_1}, a_{k_2}, ..., a_{k_\alpha})$ |
| $(b_1, b_2, ..., b_N)$ | $(b_{l_1}, b_{l_2}, ..., b_{l_\beta})$ |
| $L$ | $q-w_p$ |
| $U$ | $q-1$ |

$$X = \sum_{i=1}^{M} \sum_{j=1}^{N} g(x_i + y_j) a_i b_j = \eta_p(D)$$

$$g: \mathbb{Z} \to \mathbb{Z} \quad g(n) = \begin{cases} 1 & L \leq n \leq U \\ 0 & else \end{cases}$$

The solution of technical problem is the number of coalitions with critical player *p*. We repeat steps 3-4 for all the players obtaining the number of coalitions with particular players critical.

The Banzhaf indices of players can be calculated using formula 4:

$$\beta_p = \eta_p(G) / \sum_{i \in P} \eta_i(G)$$

**Complexity analysis:**

**1** can be done in O(n) time.

From theorem 3, **2** may be done in $O(n\alpha+n\beta)$ time, where $\alpha$ is the number of generating function $G_A(x)$ coefficients and $\beta$ the number of $G_B(x)$ coefficients. Sets A and B have at most *n*/2 players, thus generating functions $G_A(x)$ and $G_B(x)$ have at most $2^{n/2}$ non-zero coefficients (in pessimistic case each of $2^{n/2}$ coalitions has distinct weights sum). As a consequence **2** can be done in $O(n2^{n/2})$ time.

**3** and **4** are executed for each player, hence *n* times. Let *p* be a player from set A. The case when *p* is from set B is symmetrical.

According to theorem 4, the generating function $G_A(x)$ can be divided by $(1+x^{w_p})$ in $O(\alpha)$ time. The resulting generating function $G_C(x)$ has at most $\alpha$ non-zero coefficients. Thus **3** can be done in $O(2^{n/2})$.

According to theorem 1 the technical problem can be solved in O(N+M) time, where N and M are the numbers of coefficients $x_i$ and $y_i$. As a consequence **4** can be done in $O(\alpha+\beta)$ time, where $\alpha,\beta$ are the numbers of non-zero generating functions $G_A(x)$ and $G_B(x)$ coefficients. Hence **4** can be done in $O(2^{n/2}+2^{n/2})=O(2*2^{n/2})=O(2^{n/2})$ time.

Since **3** and **4** are executed *n* times the time complexity of this part of the algorithm is $O(n\alpha+n\beta)$. This can be also written as $O(n2^{n/2})$.

As a consequence the whole algorithm for calculating indices of all players has time complexity of $O(n\alpha+n\beta)=O(n2^{n/2})$.

It is worth noting that $\alpha,\beta$, the numbers of non-zero generating functions coefficients, which are equal to the number of distinct weights sums among coalitions, may be limited by quota *q* of the game. The player can't be critical in a coalition if it's weights sum without him is already greater or equal *q*. Hence the time complexity of the whole algorithm, after little modification, could be also written as $O(nq)$. Thus the algorithm keeps the pseudo-polynomial time complexity of generating functions algorithm and unlike it has pessimistic time complexity $O(n2^{n/2})$.

The algorithm described in this section is quite basic. It can be further improved. Those modifications don't change it's pessimistic computational complexity but reduce the number of operations during computations.

Modifications:

1. In point **3**, when we divide generating functions, the zero GF coefficients can be removed from the list as they appear. See also modification 2.

2. If we don't remove zero GF coefficients in point **3**, the tables with u=u(j) in technical problem are equal for all players from the sets A and B. It is sufficient to calculate *u* values only twice: for the first player in set A and first player in set B.

3. If we don't remove zero GF coefficients in point **3** the F function in technical problem (see formula 10) may be calculated only twice: for the list $K_A$ and $K_B$. This modification with modification 2 is probably better optimization than modification 1.

4. In all cases when we are computing generating functions the coefficients corresponding to weight sums greater than quota *q* may be omitted.

5. If some players share the same weight it's sufficient to calculate number of critical coalitions for one of them and use this value for all of them.

# 9 Comparison with other algorithms

| Algorithm | Pessimistic complexity | Psuedopolynomial complexity | Complexity expressed most accurately | Memory complexity | Data (type) | MVG | Notes | Info |
|---|---|---|---|---|---|---|---|---|
| Direct Enumeration | $\Theta(n2^n)$ | - | $\Theta(n2^n)$ | $\Theta(n)$ | any | Y | naive algorithm | - |
| Klinz & Woeginger "Partitioning" | $\Theta(n^2 2^{n/2})$ | - | $\Theta(n^2 2^{n/2})$ | $\Theta(2^{n/2})$ | any* | N | *int in [7] other data types possible | [7] |
| Generating Functions,"table implementation" | $\Theta(nq)$ | $\Theta(nq)$ | $\Theta(nq)$ | $\Theta(q)$ | int | Y | possible even $q>2^n$ | [1,3,4,5,6,7,**17**] |
| Generating Functions, "list implementation" | $O(n2^n)$ | $O(nq)$ | $\Theta(nC)$ | $\Theta(C)$ $O(2^n)$ | int | Y | C - number of non-zero GF coefs. $a_k x^k$, k<q $C \leq q$ the complexity in literature is O(n*n*q) but can be implemented as here with O(nq) | [1,3,4,5,8,9,17], here |
| Matsui Enumeration | ? | ? | $O(n^2 |minW|)$ | $O(n^2)$ | ? | ? | minW is family of minimal winning coalitions | [10] |
| Generating Functions "Partitioning" (algorithm from this paper) | $O(n2^{n/2})$ | $O(nq)$ | $\Theta(nC_1 + nC_2)$ | $\Theta(C_1+C_2)$ $O(2^{n/2})$ | int | N | $P = A \cup B$ $A \cap B = \emptyset$ $C_1, C_2$ - number of non-zero GF coeffs. from A and B, $a_k x^k$ for k<q $C_1 + C_2 \leq 2C$ | here |

*Table 1: The comparison of properties of exact algorithms for voting games. Optimized implementations are compared. Symbol n denotes number of players(problem size), q is the quota*

In table 1 we compare the complexities in case of calculating indices for all players. Optimized implementations are compared. For instance, in case of generating functions only coefficients corresponding to coalitions with weights sums lower than quota q are computed.

The pessimistic computational complexity is classic computational complexity expressed as a function of problem size. In our case this is the number of players.

The pseudopolynomial complexity is the polynomial complexity of the algorithm expressed as a function of problem size and the size of some number present in the instance of the problem. It is worth noting that numbers in the problem instance may exceed $2^n$.

Complexity expressed most accurately is complexity expressed as a function of problem size and some value that isn't defined explicitly in the instance of the problem. This complexity expresses the number of operations made in algorithm better than other columns.

Memory complexity is classical memory complexity of the algorithm. If two formulas are supplied, it means that one of them assesses complexity more accurately, however calculating it may be not possible before solving the instance of the problem. The second value is a direct function of the problem size and thus is easy to determine.

Data type tells whether the algorithm is applicable for types other than integers.

MVG tells if the approach in algorithm can be adapted for multiple weighted voting games. In such case complexities may differ.

# 10 Genral case (non-integer weights)

In case weights are not integers the algorithm can be used almost without modification. The algorithm no longer has pseudopolynomial complexity. The number of distinct sums of weights of coalitions no longer can be bounded by sum of all players weights (or quota). Thus the algorithm has pessimistic complexity $O(n2^{n/2})$, which is a slight improvement over Klinz and Woeginger $O(n^2 2^{n/2})$ for calculating Banzhaf indices of all players. Non-integer version of the algorithm can only be used with list implementation of generating functions algorithm (as is described in this paper).

# 11 Further research, open problems and conclusion

In this paper algorithm for calculating Banzhaf indices for all players was designed. Algorithm has pessimistic time complexity $O(n2^{n/2})$ and therefore is positive answer to the open problem stated in [2] - the existence of algorithm for calculating exactly Banzhaf indices of all players in time less than $O(n^2 2^{n/2})$. In case weights are integers, devised algorithm also shares the property of generating functions algorithm - it has pseudopolynomial time complexity $O(nq)$.

Constructed algorithm can be modified to calculate Shapley-Shubik power index (approach from [7] can be mixed with generating functions of two variables similarly as in this paper).

The new open problem is to devise algorithm to calculate exactly Banzhaf indices of all players with time complexity lower than $O(n2^{n/2})$. It would be most valuable if such algorithm could also keep the pseudopolynomial complexity in case weights are integers.

It seems that approaches from [7] and this paper cannot be used for multiple weighted voting games. Another challenging open problem is to devise algorithm that calculates Banzhaf indices of all players in multiple weighted voting game with pessimistic time complexity lower than $O(n2^n)$.

**Bartosz Meglicki**

*Former student of Warsaw University of Technology, Faculty of Mathematics and Information Science*

*meglickib@student.mini.pw.edu.pl*

*(any feedback is welcome)*